\documentclass[prl]{revtex4}
\usepackage{bm}
\usepackage{epsfig}
\usepackage{amsmath}
\begin{document}
\title{Buckled nano rod - a two state system and its dynamics}
\author{Aniruddha Chakraborty,$^{1,2}$ Sayan Bagchi $^{1}$ and K. L. Sebastian$^{1,2}$}
\affiliation{\emph{$^{1}$Department of Inorganic and Physical
Chemistry, Indian Institute of Science, Bangalore, 560012, India,}\\
\emph{$^{2}$ Jawaharlal Nehru Center for Advanced Scientific
Research, Bangalore, 560064, India.}}
\date{\today}

\begin{abstract}
\noindent We consider a suspended elastic rod under longitudinal
compression. The compression can be used to adjust potential
energy for transverse displacements from harmonic to double well
regime. The two minima in potential energy curve describe two
possible buckled states at a particular strain. Using transition
state theory (TST) we have calculated the rate of conversion from
one state to other. If the strain $\varepsilon$ is between
$\varepsilon_c$ and $4 \varepsilon_c$, the saddle point is the
straight rod. But for $\varepsilon_c < 4 \varepsilon_c$, the
saddle is S-shaped. At $\varepsilon_c = 4 \varepsilon_c$ the
simple TST rate diverges. We suggest methods to correct this
divergence, both for classical and quantum calculations. We also
find that zero point energy contributions can be quite large (as
large as $10^9$) so that single mode calculations can lead to
large errors in the rate.
\end{abstract}
\maketitle
\begin{figure}
\centering \epsfig{file=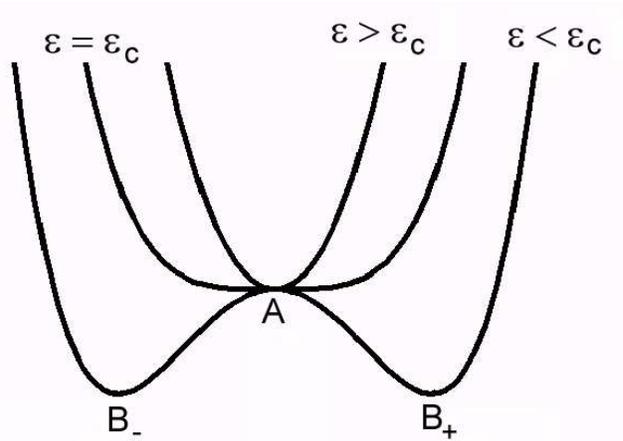,width=0.5\linewidth} \newline
\caption{Potential energy $V$ as a function of the fundamental
mode displacement $Y$. The shape of the potential energy is harmonic for $\protect%
\varepsilon >\protect\varepsilon _{c}$, quartic for $\protect\varepsilon =%
\protect\varepsilon _{c}\equiv $ critical strain ($\protect\varepsilon %
_{c}<0 $) and a double well for $\protect\varepsilon
<\protect\varepsilon _{c}$.} \label{figpotential}
\end{figure}

 \noindent Considerable attention has recently been paid
to two-state nano-mechanical systems \cite{Roukes, Craighead,
Park, Erbe, ClelandAPL,Rueckes,Blencowe} and the possibility of
observing quantum effects in them. Roukes \textit{et al.}
\cite{Roukes} proposed to use an electrostatically flexed
cantilever to explore the possibility of macroscopic quantum
tunnelling in a nano-mechanical system. Carr \textit{et al.} \cite
{Carr,ClelandAPL} suggested using the two buckled states of a
nanorod and investigated the possibility of observing quantum
effects. A suspended elastic rod of rectangular cross section
under longitudinal compression is considered. As the compressional
strain is increased to the buckling instability \cite{Carr}, the
frequency of the fundamental vibrational mode drops continuously
to zero. Beyond the instability, the system has a double well
potential for the transverse motion (see Fig. \ref{figpotential}).
The two minima in the potential energy curve describe the two
possible buckled states at that particular strain \cite{Carr} and
the system can change from one to the other, under thermal
fluctuations or quantum tunneling. We use $L$, $w$ and $d$
(satisfying $L>>w>>d$) to denote the length, width and thickness
of the rod \cite{Carr, Wybourne, CarrAPL, Lawrence}. $F$ is the
linear modulus (energy per unit length) of the rod and is related
to the elastic modulus $Q$ of the material by $F=Qwd$. The bending
moment $\kappa $ is given by $\kappa ^{2}=d^{2}/{12}$. We take the
length of the uncompressed rod to be $L_{0}$. We apply compression
on the two ends, reducing the separation between the two to $L$.
If $y(x)$ denotes the displacement of the rod in the `$d$'
direction,
\begin{figure}
\centering \epsfig{file=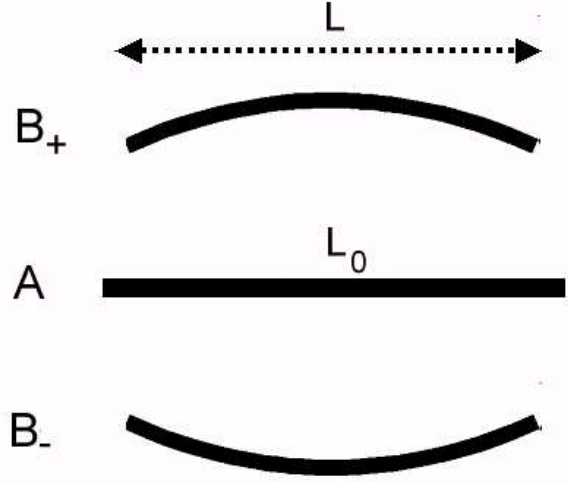,width=0.5\linewidth} \newline
\caption{The rod under compression: The central figure (A) shows
the uncompressed rod of length $L_{0}$. On compressing to length
$L$, the rod buckle, either to $B_{-}$ or to $B_{+}$.}
\label{figbuckledstate}
\end{figure}
then the length of the rod
$L_{total}=\int_{0}^{L}dx\sqrt{1+(y^{\prime })^{2}}\approx
L+1/2\int_{0}^{L}dx(y^{\prime })^{2}$. The compression causes a
contribution to the potential energy
$V_{elastic}=F/(2L_{0})(L_{total}-L_{0})^{2}$. In addition,
bending of the rod in the `$d$'
 direction cause bending energy $V_{b}=F\kappa ^{2}/{2}
\int_{0}^{L}dx(y^{\prime \prime })^{2}$. Thus the total potential
energy is given by $V[y(x)]={1}/{2}\int_{0}^{L}dx(F\kappa
^{2}(y^{\prime \prime })^{2}+F\varepsilon (y^{\prime })^{2})+{F}/(8L_{0})%
(\int_{0}^{L}dx(y^{\prime })^{2})^{2}+{F}/(2L_{0})(L-L_{0})^{2}$.
Here $\varepsilon =(L-L_{0})/L_{0}$ is the strain. Extremisation
of potential energy functional with respect to $y(x)$ leads to
\begin{equation}
F\kappa ^{2}\frac{\partial ^{4}y}{\partial x^{4}}-[F\varepsilon \frac{%
\partial ^{2}y}{\partial x^{2}}+\frac{F}{2L_{0}}(\int_{0}^{L}dx(y^{%
\prime }(x))^{2})\frac{\partial ^{2}y}{\partial x^{2}}]=0  \label{B26}
\end{equation}
\noindent and the hinged end points have boundary conditions
$y(0)=y(L)=0$ and $y^{\prime \prime }(0)=y^{\prime \prime }(L)=0$.
If $\varepsilon > \varepsilon _{c}=-\kappa ^{2}\pi ^{2}/{L^{2}}$
then, the only solution to Eq. (\ref{B26}) is $y(x)=0$ if
$\varepsilon
>\varepsilon _{c}$. But if $\varepsilon
<\varepsilon _{c}$, two bucked states are possible.
They are $y(x)=\pm A%
\sqrt{{2}/{L}}\sin ({\pi }x/{L})$, with $A=\pm \sqrt{{%
2L_{0}L^{2}}(\varepsilon _{c}-\varepsilon )/{\pi ^{2}}}$. For
$\varepsilon <\varepsilon _{c}$, all the normal modes of vibration
about these are stable. The solution $y(x)=0$ is now a saddle
point. One can calculate the barrier height for the process of
going from one buckled
state to the other over the saddle (linear geometry) as $\Delta E_{Barrier}^{Linear}={%
FL_{0}}(\varepsilon -\varepsilon _{c})^{2}/{2}$. The kinetic
energy of the rod is $\mu/ 2 \int_{0}^{L}\mu y_{t}^{2}dx$, where
$\mu ={m}/{L_{0}}$ is the mass per unit length. Using the
boundary conditions for the hinged end points, we find the normal modes of the rod, $%
y(x,t)=y_{n}(x)e^{i\omega _{n}t}$. At the saddle point, we obtain $%
y_{n}(x)=A_{n}\sqrt{{2}/{L}}\sin ({n\pi x}/{L})$, with
$n=1,2,3...$. The normal mode frequencies at the saddle point are
given by $\omega
_{Linear,n}^{\ddagger }=\omega _{0}\;n\,\sqrt{n^{2}-{\varepsilon }/{%
\varepsilon _{c}}}$, where $\omega _{0}={\pi ^{2}\kappa }/{L^{2}}\sqrt{%
{F}/{\mu }}$. $n=1$ is the unstable mode and it has the imaginary
frequency $\omega _{Linear,1}^{\ddagger }=i\Omega _{Linear}$,
where $\Omega _{Linear}=\omega _{0}\;\,\sqrt{{\varepsilon
}/{\varepsilon _{c}}-1}$. For the buckled state, the normal modes
are the same as at the saddle point, but the normal mode
frequencies are different. They are $\omega _{n}=\omega
_{0}n\sqrt{n^{2}-1}$ for $n>1$, while $\omega _{1}=\omega
_{0}\sqrt{2\left({\varepsilon }/{\varepsilon _{c}}-1\right) }$.
The rate expression using
classical TST is (Eq. 3.14 of reference \cite{Hanggi}) is $%
R_{f}^{classical}=\Omega _{Linear}\prod_{n=1}^{N}\left|{\omega
_{n}}/{\omega _{Linear,n}^{\ddagger }}\right| e^{-\Delta
E_{Barrier}^{Linear}/kT}$, where $N$ denotes the total number of
transverse modes of the rod. \ One makes a negligible error by
taking the value of $N$ to be infinity and this leads to
\begin{align}
\label{B38}
R_{f}^{classical}=\frac{\sqrt{F}}{2L\sqrt{\mu }}\sqrt{\Gamma \left( 2-\sqrt{%
\frac{\varepsilon }{\varepsilon _{c}}}\right) \Gamma \left( 2+\sqrt{\frac{%
\varepsilon }{\varepsilon _{c}}}\right) (\varepsilon
_{c}-\varepsilon )} e^{-\frac{FL_{0}}{2kT}(\varepsilon
-\varepsilon _{c})^{2}}.
\end{align}
\noindent The rate expression using quantum TST is given by
\begin{align}
\label{B39} R_{f}^{quantum}=\frac{\Omega
_{Linear}e^{-\frac{FL_{0}}{2kT}(\varepsilon
-\varepsilon _{c})^{2}}}{2\pi \sin (\frac{\hbar \Omega _{Linear}}{2kT})}%
\sinh (\frac{\hbar \omega _{1}}{2kT})\prod\limits_{n=2}^{N}\frac{\sinh (%
\frac{\hbar \omega _{n}}{2kT})}{\sinh (\frac{\hbar \omega
_{Linear,n}^{\ddagger }}{2kT})}.
\end{align}
The above rates have a problem. \noindent As $\sqrt{{\varepsilon }/{%
\varepsilon _{c}}}\rightarrow 2$, $\omega _{2}\rightarrow 0$ the
second buckling instability sets in and the rates diverge. As the
rod is compressed, first the mode $A_{1}\sqrt{{2}/{L}}\sin ({\pi
x}/{L})$ becomes unstable and this is the first buckling
instability and the rod buckles as a result of this. The length at
which this occurs shall be denoted by $L_{f}$. If one supposes
that the rod is compressed further
keeping the straight rod configuration, then at a length $L_{s}$, the mode $%
A_{2}\sqrt{{2}/{L}}\sin ({2\pi x}/{L})$ too would become unstable
and this is the second buckling instability.
\begin{figure}
\centering \epsfig{file=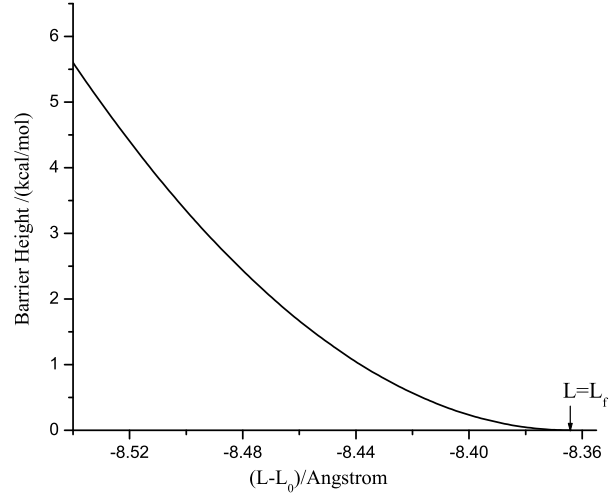,width=0.5\linewidth}
\caption{Plot of $\Delta E_{Barrier}^{Linear}$ against ($L-L_{0}$)
for a silicon rod of dimensions $L_{0}=1000 \:\AA$, $w=200 \:\AA$
and $d=100\:\AA$. For this rod the first three instabilities occur
at $L_{f}-L_{0}=-8.364\:\AA$, $L_{s}-L_{0}=-35.354\:\AA$ and
$L_{t}-L_{0}=-89.238\:\AA$ respectively. The first buckling
instability is shown by an arrow, the second and third buckling
instabilities are not shown in figure.}\label{figEact100}
\end{figure}
\begin{figure}
\centering \epsfig{file=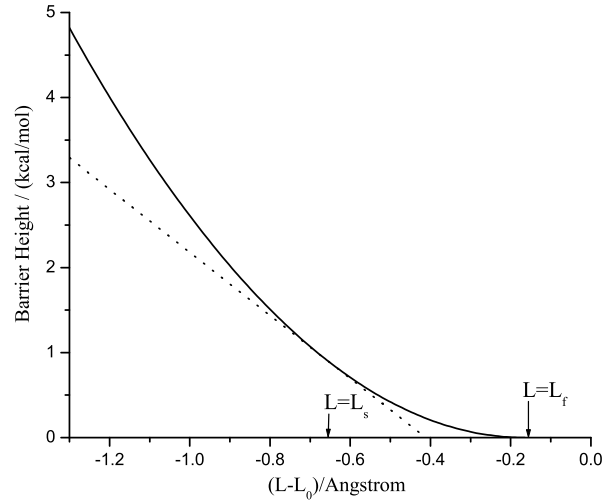,width=0.5\linewidth}
\caption{Plot of $\Delta E_{Barrier}^{Bent}$ against ($L-L_{0}$),
for a silicon rod of dimensions $L_{0}=500\:\AA $,$w=20\:\AA $,
$d=10\:\AA $. For this rod the first three instabilities occur at
$L_{f}-L_{0}=-0.1646\:\AA $, $L_{s}-L_{0}=-0.6597\:\AA $ and
$L_{t}-L_{0}=-1.4893\:\AA $ respectively and all three are shown
in figure using crosses. Solid line is for linear saddle point
(valid in the regime $L>L_{s}$ ), dashed line is for bent saddle
point (valid in the regime $L<L_{s}$).} \label{figEact50}
\end{figure}
\noindent For $\varepsilon >4\varepsilon _{c}$, there is only one
saddle point. But for $\varepsilon <4\varepsilon _{c}$, due to the
second instability, the saddle point bifurcates into two. In a
similar fashion one can have the third instability at a length
$L_{t}$ etc. In order to analyze the rate near and beyond the
second
buckling instability, we assume that the displacement has the form $%
y_{0}(x)=A_{1}\sqrt{{2}/{L}}\sin ({\pi x}/{L})+A_{2}\sqrt{{2}/{L}%
}\sin ({2\pi x}/{L})$. Using this, the elastic potential energy is
given by
\begin{align}
\label{B41} V(A_{1},A_{2})=\frac{F\pi
^{4}(A_{1}^{2}+4A_{2}^{2})^{2}}{8L^{4}L_{0}}+ \frac{F\pi
^{2}A_{1}^{2}(\varepsilon -\varepsilon _{c})}{2L^{2}}+\frac{2F\pi
^{2}A_{2}^{2}(\varepsilon -4\varepsilon _{c})}{L^{2}}.
\end{align}
Finding the extrema of this potential leads to the following three
solutions for ($A_{1}$, $A_{2}$): (a) ($0$, $0$): this is the
straight rod configuration. Between first and second buckling
(i.e. $L_{s}<L<L_{f}$ ), this is the saddle point. But after the
second buckling, it is no longer a saddle, but it becomes a hill
top. It has the energy $E_{hilltop}=0$. (b) ($\pm {2}/{\pi
}\sqrt{LL_{0}(\varepsilon _{c}-\varepsilon )}$, $0$):
These are the buckled states and both of them have the same energy $E_{b}=-%
{FL_{0}}(\varepsilon -\varepsilon _{c})^{2}/2$. (c) ($0$, $\pm
{1}/{\pi }\sqrt{LL_{0}(4\varepsilon _{c}-\varepsilon )}$): These
are the two new saddle points that arise from the bifurcation of
the one that existed for $4\varepsilon _{c}<\varepsilon $. At
these saddle points, the rod has a bent (S shaped) geometry. These
two have the same energy $E_{saddle}^{Bent}=-{FL_{0}}(\varepsilon
-4\varepsilon
_{c})^{2}/2$. Beyond the second buckling instability, the barrier height is given by $%
\Delta E_{Barrier}^{Bent}=-{3FL_{0}\varepsilon _{c}}(-2\varepsilon
+5\varepsilon _{c})/2$. Near the saddle, the normal mode frequencies are given by: $%
\omega _{Bent,1}^{\ddagger }=i\Omega _{Bent}$, $\omega
_{Bent,2}^{\ddagger }=\omega _{0}\sqrt{8({\varepsilon
}/{\varepsilon _{c}}-4)}$ and for $n>2$,
$\omega _{Bent,n}^{\ddagger }=\omega _{0}n\sqrt{n^{2}-4}$. In the above $%
\omega _{Bent,1}^{\ddagger }$ has an imaginary frequency with
$\Omega _{Bent}=\sqrt{3}\omega _{0}$. Now the classical rate
beyond the second buckling instability can be calculated taking
the saddle to be the bent configuration:
\begin{equation}
R_{s}^{classical}=4\sqrt{3}\omega _{0}\sqrt{\frac{\varepsilon
_{c}-\varepsilon }{4\varepsilon _{c}-\varepsilon }}e^{\frac{%
3FL_{0}\varepsilon _{c}}{2kT}(-2\varepsilon +5\varepsilon _{c})}.
\label{B72}
\end{equation}
\noindent It is interesting that the normal modes for this saddle retain
their stability, irrespective of what the compression is. The quantum rate
expression is given by
\begin{align}
R_{s}^{quantum}=\frac{\Omega _{Bent}e^{\frac{3FL_{0}\varepsilon _{c}}{2kT}%
(-2\varepsilon +5\varepsilon _{c})}}{\pi \sin (\frac{\hbar \Omega _{Bent}}{%
2kT})}\sinh (\frac{\hbar \omega
_{1}}{2kT})\prod\limits_{n=2}^{N}\frac{\sinh (\frac{\hbar \omega
_{n}}{2kT})}{\sinh (\frac{\hbar \omega _{Bent,n}^{\ddagger
}}{2kT})}. \label{B73}
\end{align}
\noindent Near the second buckling instability ($\sqrt{{\varepsilon }/{%
\varepsilon _{c}}}\rightarrow 2$) vanishes, causing the rates in
Eq. (\ref{B38}), Eq. (\ref{B39}), Eq. (\ref{B72}) and Eq.
(\ref{B73}) to diverge. The cure for the divergence is simple for
the classical rate and is given below in the equations
(\ref{B47a}) and (\ref{B48a}). All that one has to do is to
include the quartic term in $A_{2}$ of Eq. (\ref{B41}) in the
evaluation of partition function for the second mode at the
saddle. All the other modes (at the saddle as well as at the
reactant) are treated as harmonic. In the regime where $L>L_{s}$
the rate is then given by (saddle is the straight rod)
\begin{align}
 R_{i}^{classical}=
\frac{e^{-\frac{FL_{0}}{2kT}(\varepsilon -\varepsilon
_{c})^{2}}}{2L}f_{int}(4-\frac{\varepsilon }{\varepsilon _{c}},\frac{kT}{%
2FL_{0}\varepsilon _{c}^{2}}) \sqrt{\frac{F(-\varepsilon _{c})}{\pi \mu }(\frac{\varepsilon }{%
\varepsilon _{c}}-1)\Gamma (3-\sqrt{\frac{\varepsilon }{\varepsilon _{c}}}%
)\Gamma (3+\sqrt{\frac{\varepsilon }{\varepsilon _{c}}})},
\label{B47a}
\end{align}
\noindent where $f_{int}(a,b)=\int_{-\infty }^{\infty
}dye^{^{-ay^{2}-by^{4}}}$. In the regime where $L<L_{s}$ the rate
is given by (saddle is the bent rod)
\begin{align}
R_{i}^{classical}=\frac{e^{-\frac{FL_{0}}{2kT}(\varepsilon
-\varepsilon
_{c})^{2}}}{L}\sqrt{\frac{2F(-\varepsilon _{c})}{\mu }(\frac{\varepsilon }{%
\varepsilon _{c}}-1)}\sqrt{\frac{3}{\pi }} f_{int}(4-\frac{\varepsilon }{%
\varepsilon _{c}},\frac{kT}{2FL_{0}\varepsilon _{c}^{2}}).
\label{B48a}
\end{align}
\noindent Now we follow the work of Voth, Chandler and Miller
\cite{Voth} for rate calculation using quantum mechanical
transition state theory near the second buckling instability
(where $\omega _{Linear,2}^{\ddagger }$ is small).
\begin{figure}
\centering \epsfig{file=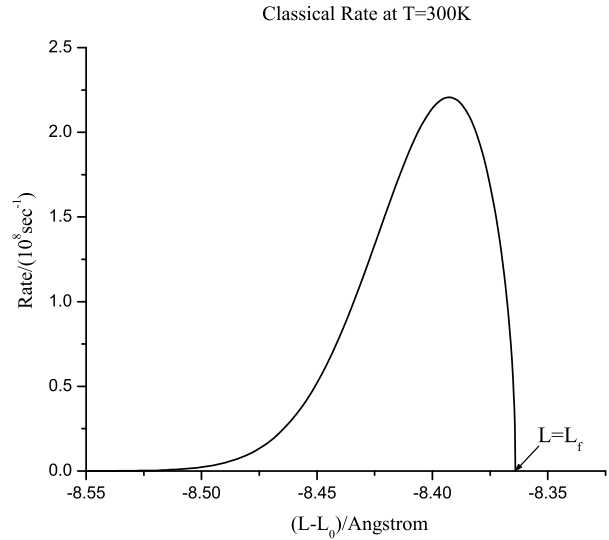,width=0.5\linewidth}
\caption{Plot of rate of crossing from one buckled state to the
other using classical transition state theory, for a silicon rod
of dimensions $L_{0}=1000 \:\AA$, $w=200 \:\AA$, $d=100 \:\AA$ at
$T=300\:K$. For this rod the first three instabilities occur at
$L_{f}-L_{0}=-8.364\:\AA$, $L_{s}-L_{0}=-35.354\:\AA$ and
$L_{t}-L_{0}=-89.238\:\AA$ respectively. The first buckling
instability is shown by an arrow, the second and third buckling
instabilities are not shown in figure.}\label{figC100T300}
\end{figure}
\begin{figure}
\centering \epsfig{file=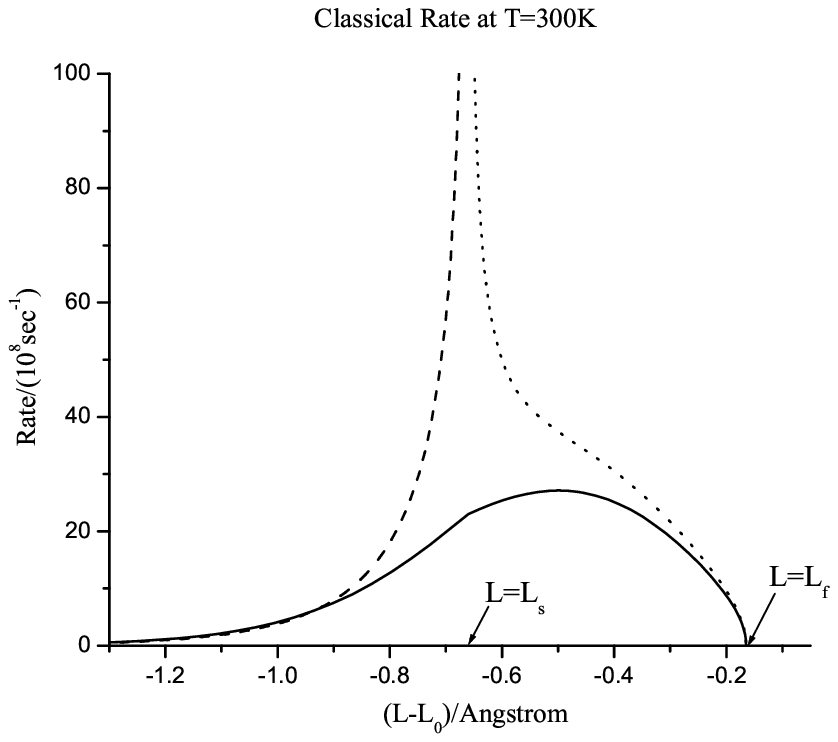,width=0.5\linewidth}
\caption{Plot of rate of crossing from one buckled state to the
other using classical TST, for a silicon rod of dimensions
$L_{0}=500\:\AA $,$w=20\:\AA $, $d=10\:\AA $ at $T=300\:K$. For
this rod the first three instabilities occur at
$L_{f}-L_{0}=-0.1646\:\AA $, $L_{s}-L_{0}=-0.6597\:\AA $ and $
L_{t}-L_{0}=-1.4893\:\AA $ respectively. Dotted line is for linear
saddle point, dashed line is for bent saddle point and solid line
include quartic terms for the second mode at saddle point.}
\label{figC50T300}
\end{figure}
\noindent The QTST rate under the harmonic approximation diverges
at temperature $T={\hbar \Omega }/({2k\pi })$, but using this
method it is possible to avoid the divergence \cite{Voth}. For
this we go beyond the harmonic approximations for the first two
modes. The Hamiltonian for the first two modes ($A_{1},A_{2}$) at
the saddle may be written as: $H={p_{1}^{2}}/({2\mu
})+{p_{2}^{2}}/({2\mu})+V(A_{1},A_{2})$, where $p_{i}=\mu
\overset{.}{A}_{i}$ is the momentum operator canonically conjugate
to $A_{i}$. The Lagrangian is $L(A_{1},A_{2})=
{\mu}/{2}(\overset{.}{A_{1}}^{2}+\overset{.}{A_{2}}^{2})-V(A_{1},A_{2})$.
The ``centroid'' partition function for these two coupled modes is
given by the path integral $Q^{\ast}=\int DA_{2}(\tau )\int
DA_{1}(\tau )\delta (\overline{A}
_{1}-A_{1}^{\ddagger })e^{-S[A_{1}(\tau ),A_{2}(\tau )]/\hbar }$. Here $%
\overline{A}_{i}$ denotes the position of the centroid, defined by $%
\overline{A}_{i}={1}/(\beta \hbar )\int_{0}^{\beta \hbar }d\tau
A_{i}(\tau )$. Note that the centroid for the first mode is constrained at $%
A_{1}^{\ddagger }$, where $A_{1}^{\ddagger }$ is the value at the
saddle \cite{Voth} and is equal to zero. We calculate the
frequency of
the unstable mode $\Omega $ and the partition function for the second mode $%
Q_{2}^{\ddagger}$ using a variational principle based on the trial
action \cite {Voth,Path2} $ S_{trial}=\int_{0}^{\beta \hbar }d\tau
\{{1}/{2}\mu {\overset{.}{A_{1}}^{2}(\tau )}-{1}/{2}\mu \Omega
^{2}A_{1}^{2}(\tau)+L_{2}(\overline{A}_{2})+{1}/{2}\mu {\overset{.}{A_{2}}^{2}(\tau )}+{1}/{2}%
\mu \omega _{2}^{2}(\overline{A}_{2})(A_{2}(\tau
)-\overline{A}_{2})^{2}\}$. One can determine $\Omega ^{2}$, $\omega _{2}^{2}(\overline{A}%
_{2}) $ and $L_{2}(\overline{A}_{2})$ variationally, so as to get
the best possible value for $Q^{\ast }$. Once these values are
obtained, we can proceed to calculate the rate, because in the
trial action, the two modes are decoupled. Now the partition
function of the second vibrational mode at the saddle $Q_{2}^{\ast
}$ may be approximated by $Q_{2}^{\ast }={\sqrt{({\mu kT})/{2\pi
\hbar ^{2}}}}\int_{-\infty }^{\infty }d\overline{A}_{2}e^{-\frac{%
W_{2}(\overline{A}_{2})}{kT}}$, with an effective potential $W_{2}(%
\overline{A}_{2})=kT\log [{2kT}/(\hbar
\omega_2(\overline{A}_2))\sinh({\hbar \omega
_{2}(\overline{A}_{2})}/({2kT}))]+L_{2}(%
\overline{A}_{2})$. The tunneling current for the first mode may
be  taken as ${\Omega}/( {2\pi \sin (\frac{\hbar \Omega
}{2kT})})$, where $\Omega$ is the variationally determined
frequency of the reactive mode. In the regime where $L>L_{s}$ the
rate may be calculated using (transition state is assumed to be
straight rod)
\begin{align}
R_{i}^{quantum}=\frac{\Omega e^{-\frac{FL_{0}}{2kT}(\varepsilon
-\varepsilon _{c})^{2}}}{\pi \sin (\frac{\hbar \Omega
}{2kT})}\sinh (\frac{\hbar \omega _{1}}{2kT})\sinh (\frac{\hbar
\omega _{2}}{2kT})Q_{2}^{\ast
} \prod\limits_{n=3}^{N}\frac{\sinh (\frac{\hbar \omega _{n}}{2kT})}{\sinh (%
\frac{\hbar \omega _{Linear,n}^{\ddagger }}{2kT})}.
\end{align}
\noindent In the regime where $L<L_{s}$ the rate is given by (transition
state is assumed to be bent rod)
\begin{align}
R_{i}^{quantum}=\frac{\Omega e^{-\frac{FL_{0}}{2kT}(\varepsilon
-\varepsilon _{c})^{2}}}{\pi \sin (\frac{\hbar \Omega
}{2kT})}\sinh (\frac{\hbar \omega _{1}}{2kT})\sinh (\frac{\hbar
\omega _{2}}{2kT})Q_{2}^{\ast
}\prod\limits_{n=3}^{N}\frac{\sinh (\frac{\hbar \omega _{n}}{2kT})}{\sinh (%
\frac{\hbar \omega _{Bent,n}^{\ddagger }}{2kT})}.
\end{align}
Now we consider the Silicon rods of different dimensions, as
summarized in Table. \ref{Tableresult}. Si has an Young's modulus
$Q=130\:GPa$ and density $\rho =2230\:kg.m^{-3}$. First we
consider a rod of dimensions $100\:nm\times 20\:nm\times 10\:nm$,
considered by Carr \textit{et al} \cite{Carr}. In Fig.
\ref{figEact100} we plot the activation energy against the
compression. At $L=L_{f}$, the activation energy is zero and as
one compresses the rod, it increases rapidly, as it is
proportional to $\left( \varepsilon -\varepsilon _{c}\right)
^{2}$. Over a very short range of compression ($\sim 0.14 \:\AA$)
it increases to about $6\:kcal/mol$. The classical rate is plotted
in Fig. \ref{figC100T300}. At $L=L_{f}$, the potential is very
flat and near this $L$ the pre-factor (\textit{i.e.}, the attempt
frequency) in the rate expression vanishes. This is the reason why
the rate goes to zero as $L\rightarrow L_{f}$. It is found that
the rate increases at first as one compresses. This is due to the
increase in the pre-factor for the rate from zero. Then the rate
decreases due to the increase in the barrier height. For the
quantum calculation, we choose the value of $N$ to be equal to the
transverse degrees of freedom that would be there if one
considered an atomistic model for the rod. Thus, for this rod, we
took $N=425$. We also report (see Table \ref{Tableresult}) the
quantum enhancement factor, defined by $\Gamma $=Quantum
Rate/Classical Rate.
\begin{figure}
\centering \epsfig{file=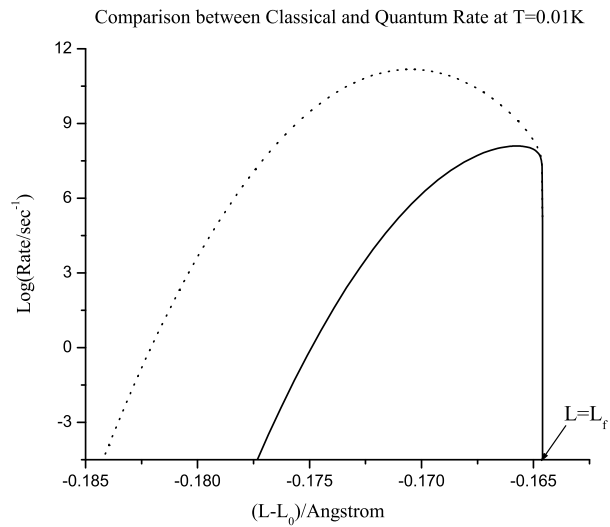,width=0.5\linewidth}
\caption{Plot of logarithm of rate of crossing from one buckled
state to the other using both classical TST and quantum TST, for a
silicon rod of dimensions $L_{0}=500\:\AA $,$w=20\:\AA $,
$d=10\:\AA $ at $T=0.01\:K$. The solid line is the result using
classical TST, dashed line is the result using quantum TST.}
\label{figCom50T001}
\end{figure}
\begin{figure}
\centering \epsfig{file=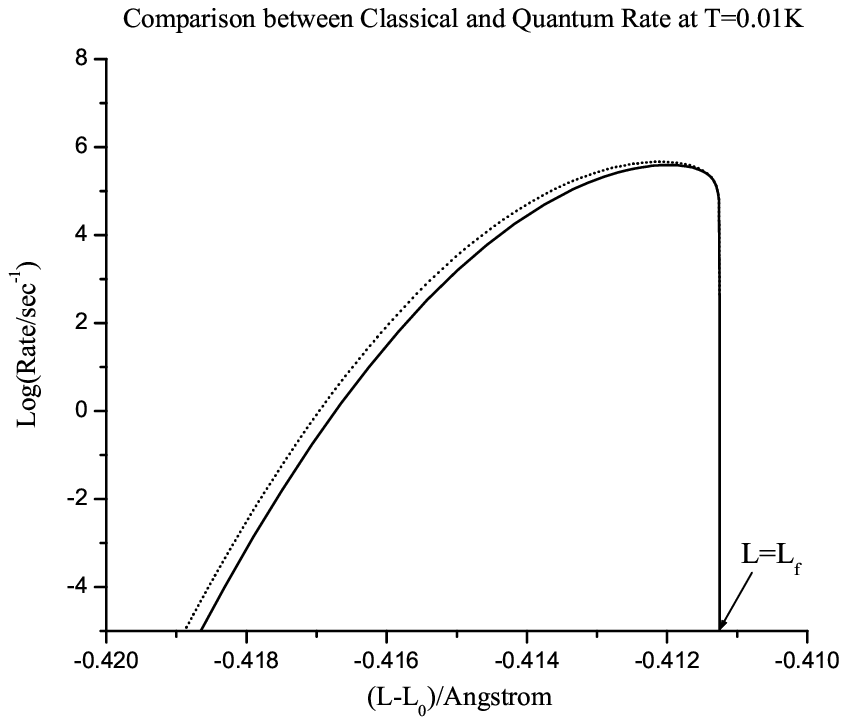,width=0.5\linewidth}
\caption{Plot of logarithm of rate of crossing from one buckled
state to the other using both classical TST and quantum TST, for a
silicon rod of dimensions $L_{0}=20000 \:\AA$, $w=200 \:\AA$ and
$d=100\:\AA$ at $T=0.01\:K$. The solid line is the result using
classical TST, dashed line is the result using quantum TST.}
\label{figCom2000T001}
\end{figure}
\noindent The values of $\Gamma$ was found to be $\approx 1$
implying that quantum effects are not important for this rod,
within the observable range of compression. For a rod of
dimensions $2000\:nm\times 20\:nm\times10\:nm$, plots of the rates
at $0.01\:K$ is given in Fig. \ref{figCom2000T001} and it shows
that quantum effects lead to an increase in the rate by about a
factor of $10$. However, the overall rate rapidly decreases to
very low values as one compresses the rod by about $0.01\:\AA$
(see the Fig. \ref{figCom2000T001}) and hence it would be very
difficult to observe this quantum enhancement experimentally. We
have also performed calculations for a hypothetical rod of
dimensions $50\:nm\times 2\:nm\times 1\:nm$. It is yet not
possible yet to have $Si$ rod of these dimensions. However it
should be possible to synthesize molecular rods of these
dimensions \cite{Schwab}. In Fig. \ref{figEact50} we have plotted
barrier height as a function of compression. In the regime $L\geq
L_{s}$, the transition state has straight rod configuration and in
the regime $L<L_{s}$, the transition state has bent configuration.
Fig. \ref{figC50T300} shows the classical rate against
compression, made at a temperature of $300\:K$. The rate obtained
using the linear transition state is seen to diverge at the second
buckling instability, but is finite for all $L>L_{s}$. Similarly,
beyond the second buckling instability the rate is calculated
using the bent saddle. Close to the instability, this rate too
diverges, but a well behaved rate can be calculated using the
approach of Voth {\it et al.} \cite{Voth} outlined above. In the
quantum rate calculation we have taken contributions from $213$
normal modes of this rod. We have compared the quantum rate with
the classical rate at $0.01\:K$ (Fig. \ref{figCom50T001}). There
is a quantum enhancement in the rate of roughly $10^{6}$. This
occurs as one compresses the rod by about $0.015\:\AA$. Again,
fabricating such a rod and doing an experiment under such
conditions is a formidable challenge. We now ask, what is the
origin of a high quantum enhancement factor in some of the cases?
This is not due to tunneling of the reactive mode
\cite{Carr,Blencowe}, but is a zero point energy effect. The
earlier analysis took only the reactive mode in their calculations
\cite{Carr,Wybourne,CarrAPL,Lawrence}, but our analysis takes all
the modes. This leads surprisingly to a decrease in the effective
activation energy. The effective quantum mechanical barrier height
may be written as $\Delta E_{Barrier}^{Linear}+\Delta E_{zero}$,
where $\Delta E_{zero}$ is the difference in zero point energy of
the modes between the buckled state and the transition state. For
the linear transition state $\Delta E_{zero}=\sum_{n=2}^{N}\hbar
\omega _{0}n\left( \sqrt{n^{2}-\varepsilon /\varepsilon _{c}}-
\sqrt{n^{2}-1}\right) $ $\cong $ $\sum_{n=2}^{N}$ $\hbar
\omega_{0}(N-1)\left( \varepsilon /\varepsilon _{c}-1\right) /2$. Notice that $%
\Delta E_{zero}$ is negative and hence leads to a lowering of the
barrier height, which is proportional to $N$. This is the reason
why the quantum effect is more pronounced for the $2000\:nm$ bar
(provided $w$ and $d$ are the same) as may be seen in Table
\ref{Tableresult}. The only possibility for observing quantum
effects seems to be exciting the rod to a higher level in the
buckled potential, suggested by Blencowe \cite{Blencowe}. Thus,
for the $2000\:nm$ bar if one keeps $L-L_0=-0.05\:nm$, the barrier
height has the value $0.0935\:kcal/mol$. The classical rate at
$T=0.01\:K$ is $1.55\times 10^{-2038}\:sec^{-1}$. The Quantum
enhancemenr factor $\Gamma=1.32\times10^{9}$. Even with this
enhancement, the net rate is far too small to be observed. But
following the idea of Blencowe \cite{Blencowe}, the system which
is initially in thermal equilibrium near the bottom of the well,
can be excited to an energy level below the barrier maximum, then
it will be more likely to tunnel through the barrier than being
thermally activated over it. In such a case the rate analysis
using only the reactive mode will be underestimating the rate by a
factor of roughly $10^{9}$!

\begin{widetext}
\begin{table}
\centering
\begin{tabular}{|c|c|c|c|c|} \hline
\multicolumn{1}{|c|}{Dimensions}  &
\multicolumn{2}{|c}{Temperature} & \multicolumn{1}{|c|}{Range of
compression over which classical} & \multicolumn{1}{|c|}{$\Gamma$}
\\
\multicolumn{1}{|c|}{}  & \multicolumn{2}{|c}{} &
\multicolumn{1}{|c|}{rate varies between $10^5$ and $10^9
\:sec^{-1}$} & \multicolumn{1}{|c|}{}
\\
\hline \multicolumn{1}{|c}{L=1000\:\AA } &\multicolumn{2}{|c}{} &
\multicolumn{1}{|c|}{}& \multicolumn{1}{|c|}{}
\\
\multicolumn{1}{|c}{w=200\:\AA} &\multicolumn{2}{|c}{300\:K} &
\multicolumn{1}{|c|}{0.14\:\AA}& \multicolumn{1}{|c|}{$\approx 1$}
\\
\multicolumn{1}{|c}{d=100\:\AA} &\multicolumn{2}{|c}{0.01K} &
\multicolumn{1}{|c|}{0.001\:\AA}& \multicolumn{1}{|c|}{$\approx
1$}
\\
\hline \multicolumn{1}{|c}{L=20000\:\AA} &\multicolumn{2}{|c}{} &
\multicolumn{1}{|c|}{}& \multicolumn{1}{|c|}{}
\\
\multicolumn{1}{|c}{w=200\:\AA} &\multicolumn{2}{|c}{300\:K} &
\multicolumn{1}{|c|}{0.6\:\AA}& \multicolumn{1}{|c|}{$\approx 1$} \\
\multicolumn{1}{|c}{d=100\:\AA} &\multicolumn{2}{|c}{0.01K} &
\multicolumn{1}{|c|}{0.006\:\AA}& \multicolumn{1}{|c|}{$\approx
10$}
\\\hline
\multicolumn{1}{|c}{L=500\:\AA} &\multicolumn{2}{|c}{} &
\multicolumn{1}{|c|}{}& \multicolumn{1}{|c|}{}
\\
\multicolumn{1}{|c}{w=20\:\AA} &\multicolumn{2}{|c}{300\:K} &
\multicolumn{1}{|c|}{1\:\AA}&
\multicolumn{1}{|c|}{$\approx 1$}\\
\multicolumn{1}{|c}{d=10\:\AA} &\multicolumn{2}{|c}{$0.01\:K$} &
\multicolumn{1}{|c|}{0.015\:\AA}& \multicolumn{1}{|c|}{$\approx
10^6$}
\\\hline
\end{tabular}
\caption{Summary of results.} \label{Tableresult}
\end{table}
\end{widetext}

\end{document}